\documentstyle[11pt,aaspp4]{article}
\input epsfig.sty
\begin{document}

\newcommand\X{{$X = N{\rm(H_2)}/W_{\rm CO}$ }}
\newcommand\ssrev{{Space Science Reviews}}

\title{ EGRET Observations of the Diffuse Gamma-Ray Emission in 
Orion:  Analysis Through Cycle 6}

\author{S. W. Digel,$^1$ E. Aprile,$^2$ S. D. Hunter,$^3$ R. Mukherjee,$^4$ 
and F. Xu$^2$}
\bigskip
\affil{1. USRA, NASA/Goddard Space Flight Center, Code 660.2, 
Greenbelt, MD 20771}
\affil{2. Physics Dept. \& Columbia Astrophysics Lab., Columbia University, New York, NY~10027}
\affil{3. NASA/Goddard Space Flight Center, Code 661, Greenbelt, MD 20771}
\affil{4. Dept. of Physics \& Astronomy, Barnard College, Columbia University, New York, NY 10027}

\begin{abstract}

We present a study of the high-energy diffuse emission observed toward
Orion by the Energetic Gamma-Ray Experiment Telescope (EGRET) on the
{\sl Compton Gamma-Ray Observatory}.  The total exposure by EGRET
in this region has
increased by more than a factor of two since a previous study.  A simple
model for the diffuse emission adequately fits the data; no
significant point sources are detected in the region studied
($l = 195^\circ$ to $220^\circ$ and $b = -25^\circ$  to $-10^\circ$) in
either the composite dataset or in two separate groups of EGRET viewing
periods considered. 
The gamma-ray emissivity in Orion is found to be
$(1.65 \pm 0.11) \times 10^{-26}$ s$^{-1}$ sr$^{-1}$ for $E >$ 100 MeV,
and the
differential emissivity is well-described as a combination of
contributions from cosmic-ray electrons and protons with approximately the
local density.  The molecular mass calibrating ratio is $N({\rm H_2})/W_{\rm
CO}$ = $(1.35 \pm 0.15) \times 10^{20}$ cm$^{-2}$ (K km s$^{-1}$)$^{-1}$.  
\end{abstract}


\section{Introduction}

This paper presents an analysis of the most extensive data set to date 
of the Orion region 
obtained by the Energetic Gamma Ray Experiment Telescope (EGRET) on the 
{\sl Compton Gamma Ray Observatory} (CGRO). 
Study of the diffuse high energy ($E>30$ MeV) gamma-ray emission 
from nearby, massive interstellar clouds permits testing
the mechanisms of gamma-ray production and measuring the 
local cosmic-ray (CR) density as well as properties of the interstellar
medium (ISM). 
The goals of this work are to 
determine the high-energy CR density in Orion, the molecular 
mass calibrating factor $X \equiv N({\rm H_2})/W_{\rm CO}$, 
and to identify any point sources or resolved variations in CR density or
$X$ within the Orion AB-Mon R2 complex of interstellar clouds.
The Orion region was previously studied by Digel, Hunter, \& Mukherjee (1995; 
hereafter DHM) using EGRET data through 1993 of CGRO.  
Since the time of the earlier work by DHM, 
the overall exposure of EGRET toward the Orion region has 
increased by more than a factor of two, and has become much more uniform.  
As EGRET is now nearing the end of the life of its spark 
chamber gas, the currently-available observations 
represent essentially all the exposure that EGRET will obtain 
toward Orion.

Most of the motivations for studying the diffuse gamma-ray 
emission in Orion remain the same. 
The interstellar clouds in Orion, comprising the Orion A and B clouds 
and the more distant Mon R2 cloud, are the nearest giant molecular 
clouds ($\sim 500$ pc), with a mass $\sim 4 \times 10^5$ M$_\odot$ 
(Maddelena et al. 1986). The clouds are well-resolved by EGRET, 
and are far from the plane in the 
outer Galaxy, so their gamma-ray emission can be studied 
essentially in isolation from the general diffuse emission of 
the Milky Way. 

Since the time of the previous work, 
more conservative cuts on zenith angle to reject earth albedo 
gamma rays were adopted for the standard EGRET data products.  This
decreases the number of photons, and the exposure, somewhat for each
viewing period.  We investigate here whether this change alters the 
findings of DHM.

To verify the production mechanisms of gamma rays in interstellar clouds, 
and to determine the CR densities for individual
molecular 
clouds, independent measurements of the interstellar
matter distribution are required.
The 
atomic hydrogen phase of the ISM is observable via the
characteristic 
21 cm line radiation. However, molecular hydrogen $\rm H_2$ 
generally cannot be
directly 
observed at interstellar conditions. The standard tracer of the large-scale distribution of $\rm H_2$ is
the 
$J = 1-0$ line of $\rm CO$ at 115 GHz. $\rm CO$
is 
the second most abundant interstellar molecule after $\rm H_2$, tends
to form under the same conditions, and is excited to the $J = 1$
rotational state by collisions with $\rm H_2$.
The relation between $N{\rm(H_2)}$ and $W_{\rm CO}$,
the integrated intensity of the CO line, is empirically known to be
approximately a proportionality; 
the proportionality constant $N{\rm(H_2)}/W_{\rm CO}$ is denoted $X$.
All determinations of $X$ require an indirect tracer of $N{\rm(H_2)}$. Helium 
and heavier elements are assumed to be distributed like the hydrogen, as
is commonly done in studies of diffuse gamma-ray emission;  the
emissivities referred to in other sections of this paper are therefore
the effective rates per hydrogen atom.  Owing to the penetration
of clouds by high-energy CRs, and the transparency of the ISM to
gamma rays, gamma-ray intensity can be used as such a tracer.
Here we use the gamma-ray emission from Orion to calibrate the
$X$-ratio and thereby infer the CR densities in the Orion 
neighbourhood. 

The recent reports of an extended region of carbon and oxygen nuclear
line emission in Orion from the COMPTEL instrument on CGRO (Bloemen et al.
1997) are another motivation for an updated study of the gamma-ray 
observations by EGRET.  To explain the observed flux of the 
$^{12}$\rm C$^*$ and $^{16}$\rm O$^*$ lines at 4.44
MeV and 6.13 MeV, respectively, a large enhancement of low-energy 
($<100$ MeV/Nucleon) CRs is
needed.  Note, however, that a recent re-evaluation of the COMPTEL background 
has shown that the Orion result was largely spurious (Bloemen et al. 1999).  
The CR enhancement factor depends on the
amount of interstellar gas
and hence on the $X$-ratio determined from EGRET analysis. 

\section{Data}

\subsection{ H I and CO}

We use the same 21 cm H I and 2.6 mm CO maps as DHM.  
Briefly, the H I surveys of Weaver \& Williams (1973) and 
Heiles \& Habing (1974) were combined and column densities 
$N\rm (H I)$ were derived on the assumption of a uniform spin 
temperature of 125 K.  The CO surveys of Maddalena et al. 
(1986) and Huang et al. (1985), as combined in Dame et al. 
(1987), were used to derive the map of integrated intensity in 
the CO line, $W_{\rm CO}$.  
The region of 
interest for the present study is $l = 195^\circ$ to 
$220^\circ$ and 
$b = -25^\circ$  to $-10^\circ$, although a $15^\circ$ wide border 
surrounding this area was also included in the CO and H I 
datasets to permit convolution with the broad PSFs (point spread functions) 
of EGRET 
in the central region.  In directions where no CO data are 
available, principally $b < -25^\circ$, we assume $W_{\rm CO} = 0$.  For 
both H I and CO emissions, only one spectral line is evident along 
lines of sight in the region of interest; although the 21 cm 
line emission in particular has broad tails in velocity, all 
of the interstellar gas along the line of sight is assumed to 
be associated with Orion (i.e., have the same density of CRs) 
in the analysis below. 
The $N\rm (H I)$ and $W_{\rm CO}$ maps were produced on the 
same grid used for binning the gamma-ray photons.

\subsection{Gamma-Ray}

We combine the data (photon counts and exposure maps) 
from all EGRET viewing periods with exposure within the 
region of interest ($l = 195^\circ$ to $220^\circ$, $b=-25^\circ$ to
$ -10^\circ$; Table 1) and a $15^\circ$ border around this region.  
Only the area within $30^\circ$ of the pointing direction for any 
given viewing period was included in the combined datasets. The 
sensitivity of EGRET for inclination angles greater than $30^\circ$ is 
relatively poor, so few photons and little exposure are lost.  
The advantage of making this truncation is that the 
relatively broad PSF far off axis (Thompson et al. 1993) need 
not be considered in the analysis; for each energy range 
only a single effective PSF is needed for the entire dataset. The gamma-ray 
data are binned on a $30^\prime$ grid in Galactic coordinates for this 
analysis. Details of the instrument design are discussed in Hughes et al. 
(1980) andKanbach et al. (1988), and the preflight and the postflight 
calibrations 
are described by Thompson et al. (1993) and Esposito et al. (1999).

In the analysis described below, the EGRET data are 
analyzed for six broad energy ranges spanning 30-10,000 MeV, 
and two integral ranges (energy $E > 100$ MeV and $E > 300$ MeV).  For 
each range, the corresponding exposure maps were derived on 
the assumption of an $E^{-2.1}$ input spectrum.
The intensity maps (photon counts divided by 
exposure) for viewing periods with large overlaps and good exposure
in the region of interest were 
intercompared to check their relative intensity calibrations.  The
seven viewing periods (1.0, 2.1, 337.0, 413.0, 419.5, 420.0, and 616.1)
were compared on just four broad energy ranges (30-100, 100-300, 300-1000,
and 1000-10,000 MeV) to improve the statistics of the comparisons.  The
relative calibrations were in good agreement except for viewing
periods 2.1 and 616.1, which were found to be significantly brighter
and fainter than the average, respectively.  The correction factors
were largest for viewing period (VP 616.1), 
ranging up to 1.9 on 30-100 MeV.  For this
late VP, the performance of the spark chamber had degraded significantly;
EGRET was operated in the narrow field of view mode during this VP, and so
the overall effect on the composite data set is small.  

Table 1 lists the numbers
of photons and mean exposure (scaled as described above) for 
the representative energy ranges $E > 100$ MeV and $E > 300$ MeV.  
The correction
factors described above have been incorporated into Table 1.
The viewing periods listed in the table are grouped by 
observation date to show the two subsets that were 
considered below to check consistency with the analysis of DHM and
to search for flaring point-source emission that might
have been more significant in the viewing periods obtained since
that work.
The total number of photons for $E > 100$ MeV in the region of 
interest is 10,257, compared to 5266 photons in DHM.  (With the 
more restrictive cuts used here for Earth albedo gamma rays, 
the previous total for the same viewing periods is 4879.)  The overall 
mean exposure has increased from $5.9\times 10^8$ cm$^2$ s (before the
more restrictive zenith angle cuts were adopted) to 
$13.5 \times 10^8$ cm$^2$ s. 

\section{Analysis}

We use the same model as DHM for the 
emission in Orion, one that has been applied in several studies 
of diffuse gamma rays dating from the work of Lebrun 
et al. (1982).  Under the assumption that high-energy CR 
electrons and protons uniformly penetrate the atomic and 
molecular gas in Orion, with the same densities
in both phases, the distribution of photon counts 
may be modeled as a linear combination of 
the $N\rm(H I)$ and $W_{\rm CO}$ maps.  In principle, allowance must also
be made for inverse-Compton emission and gamma-ray production 
on ionized gas. If the CR density were uniform, the distribution of 
gamma-ray photons observed for some energy range could be written as 
\begin{equation}
\Theta(l,b) = A N{\rm (H\ I)}_c + B W{\rm (CO)}_c + CN({\rm H\ II})_c + 
IC_c + \Sigma(D_i \delta_{l_i b_i}) + F \epsilon_c.  
\end{equation}
Here we have taken the finite spatial resolution of EGRET 
into account by convolving
the predicted distribution of gamma-ray photons with the effective PSF of
EGRET for the corresponding energy range. The subscript $c$ indicates
multiplication by the exposure map and convolution with the effective PSF,
as explained in DHM. 
$\epsilon_c$ is the exposure map itself convolved with the effective PSF for
that energy range.
In Eqn. (1), the coefficient $A$ is the emissivity 
(photons s$^{-1}$ sr$^{-1}$) per hydrogen atom, $B = 2 A X$, 
where $X = N{\rm(H_2)}/W_{\rm CO}$, $C$ is the emissivity of the ionized 
hydrogen, $D_i$ are the numbers of photons from each point source for that
energy range,
and $F$ is the isotropic diffuse emission. For the Orion 
region, the IC emission and contributions from CR interactions with ionized 
hydrogen are expected to contribute at only the several percent level, and 
largely in a smooth way that can be subsumed with the isotropic emission 
(DHM). 
Under these assumptions and approximations Eqn. (1) may be re-written as 
\begin{equation}
\Theta(l,b) = A N{\rm (H\ I)}_c + B W{\rm (CO)}_c + C \epsilon_c +
\Sigma(D_i \delta_{l_i b_i})
\end{equation}

We use the maximum likelihood method (Mattox et al. 1996) 
to fit the model (Eqn. 2) to the binned 
photon data for the various energy ranges and combinations of viewing 
periods considered. The likelihood value, $L,$ for a given set of
parameter values is the product of the probabilities that the
measured photon counts are consistent with the predicted counts in each
pixel on the assumption of Poisson statistics.
The probability of one 
model with likelihood $L_1$ better representing the data than another 
model with likelihood $L_2$ is determined from twice the difference of the 
logarithms of the likelihoods, $2(\ln L_2-\ln L_1).$ This difference, 
referred to as the test statistic $TS$, is distributed like $\chi^2$ in
the null hypothesis, with 
the number of degrees of freedom being the difference
between the number of free 
parameters in the two models.
To fit the model to the observations, the positions and fluxes of the 
point sources and the values of the other coefficients in Eqn. (2) are 
adjusted to maximize the likelihood.  
The strength of the dependence of the likelihood function
on the various parameters of the model permits their 
uncertainties or significances to be 
estimated quantitatively. Values for the different coefficients 
in Eqn. (1) can be determined separately as long as their 
corresponding maps are distinguishable, i.e., are 
linearly independent.  Figure 1 shows that the maps of $N{\rm (H I)}_c$, 
$W{\rm (CO)}_c$, and $\epsilon_c$ for the Orion region are 
distinguishable, for the representative energy range $E > 100 $
MeV.  The EGRET exposure to the region modeled is not uniform, having 
a gradient with longitude, so the sensitivity decreases at higher longitudes 
and lower latitudes. The exposure variations are accounted for 
in the maximum likelihood analysis, in the sense that regions
with greater exposure effectively have greater weight.

We first used the maximum likelihood method to 
systematically search for gamma-ray point sources in the 
region of interest.  
The point 
source search entails determining the maximum likelihood 
values of the parameters in Eqn. (2) for an assumed point 
source at each position in the $30^\prime$ grid, generating a map of 
$TS$.  Owing to the strong energy 
dependence of the effective PSF of EGRET, in application this 
test is most sensitive if $TS$ maps for subranges of energy are 
generated separately, then added together (Mattox et al. 
1996). Mattox et al. (1996) show that for the case 
of six combined maps, the values of $TS$ for the point source 
search are distributed as $\chi^2$ with 8 degrees of freedom in 
the null hypothesis.  

\section{ Results}

A maximum likelihood search for point sources was made 
for the two groups of viewing periods identified in Table 1, 
as well as for the sum of all viewing periods. No
significant source detections, greater than 4-$\sigma$ statistical
significance, were found in any of the groups of viewing periods or for
any of the energy ranges analyzed. Figure 2 shows the 
sum of the $TS$ maps for the 30-100, 100-150, 150-300, 300-500, 
500-1000, and 1000-10,000 MeV ranges for the combined set of all 
viewing periods.  The peak value of 31.2, 
near ($195.0^\circ$, $-19.5^\circ$), corresponds to a significance of 
$3.8 \sigma$. 

The extended
feature associated with this peak (Fig. 2) largely originates
in the 30-100 MeV TS map.  In this low energy range, the effective PSF of
EGRET is quite broad and the feature could represent the presence of a
soft source or
sources just outside the region of interest.  In fact, the Third EGRET
catalog (3EG) (Hartman et al. 1999) includes two point sources just below the
lower longitude limit:  3EG J0459+0544 ($193.99^\circ$, $-21.66^\circ$)
and 3EG J0530+1323 ($191.50^\circ$, $-11.09^\circ$).  

The extended excess near ($215^\circ$,$-19^\circ$) in Figure 2 has a peak
TS value of 24.0, corresponding to a statistical significance of 
3.0 $\sigma$.  However, the overall significance of the extended excess is
less; the integrated residual
intensity for $E > 100$ MeV (Fig. 3$c$) corresponds to approximately only
25 photons in a region where 280 are expected.  
The 3EG catalog (Hartman et
al. 1999) contains no sources consistent with this extended region, and a
search of NED reveals
no likely counterparts at other wavelengths.
We note that position of this excess is consistent with two sources
in the Second EGRET catalog (Thompson et al. 1995),
GRO J0545-1156 and GRO J0552-1028. 
However, both of these sources were below threshold in the 3EG 
catalog analysis.

In their earlier analysis of the same region that we model here, DHM
reported the detection of 
three marginal sources at a statistical significance of $\sim 3\sigma$. Two of 
these sources were probably the same as the unidentified sources 
GRO 0605-09 ($l=216.6^\circ$, $b=-14.4^\circ$) and 
GRO 0546-02 ($l=207.6^\circ$, $b=-15.6^\circ$) listed in the First EGRET 
catalog (Fichtel et al. 1994). The third source at $l=203.0^\circ$, 
$b=-17.5^\circ$ had not been detected by EGRET previously. None of the above 
three sources were detected in the current analysis of the complete data set. 

Since we are primarily interested in the diffuse gamma-ray emission here,
we investigated how the maximum 
likelihood values of the parameters $A, B,$ and $C$ that describe the
diffuse emission (Eqn. 2) depended on the number and positions of point
sources in the model.  The cases investigated included the following: 
no point sources,
a hypothetical source at ($215^\circ$, $-19^\circ$), and the two 3EG
sources mentioned above.
For all of these cases, the maximum likelihood values of the diffuse
parameters were the same within 1-$\sigma$,
except in the 30-100 MeV
range, where the inclusion of the two 3EG sources improved the fit
and changed the best-fitting $B$ and $C$ terms substantially.  Although
these sources were not detected with strong significance, which is not
unexpected as they are outside the region we model, we included them
in the model that we adopted for all of the analysis described below.

Figure 3 shows the EGRET observations, the maximum 
likelihood best-fitting model, and the residual map for the $E > 100$ 
MeV energy range using all of the EGRET data for Orion.
The gamma-ray intensities were obtained by dividing the photon map used
to fit the parameters by the exposure map. The good agreement between the
observed intensities and the model for $E>100$ MeV, calculated using the 
parameters for the combined groups in Table 2, is demonstrated in the
residual map shown in Figure 3$c$. The figure shows that the model intensity 
map clearly has no large-scale deviation from the observed intensity. 
Owing to the
inclusion of the two low-longitude sources, the residual intensities
near longitude 195$^\circ$ are small.
The remaining extended residual was discussed above.

Figure 3 also shows that there is no significant deviation 
from the fit in the region of Mon R2. This indicates that although Mon R2 
is 300 pc more distant than Orion (Maddalena et al. 1986), its properties
can be considered to be 
the same as those of the clouds in Orion. Further, we note that 
the residual intensity shown in Figure 3$c$ is not correlated with 
$W_{\rm CO}$, $N({\rm H\ I})$, or the total column density of interstellar 
gas. The absence of the correlation with $W_{\rm CO}$ is consistent with the 
$X$-ratio being independent of $W_{\rm CO}$ and of the position in Orion. 
There is no statistically significant variation 
of the $X$-ratio and emissivity between Orion A, Orion B, and Mon R2 
molecular clouds. The 
lack of correlation between the residual intensity and the interstellar gas is 
consistent with the assumption that the atomic and molecular gas is 
uniformly permeated by CRs, and that the CR density is 
uniform. The model (Eqn. 2) therefore provides an adequate 
description of the high-energy gamma-ray emission from the Orion region. 

The parameter values for the model fits to the combined 
EGRET data and their uncertainties
are listed in Table 2. No significant 
differences are seen from the results of DHM.  The 
uncertainties in the model parameter values have decreased as 
expected owing to the greatly improved exposure.
Interpretations of the values of the parameters are discussed in the 
following paragraphs.

Figure 4$a$ shows the differential $\gamma$-ray emissivity derived from 
the coefficient $A$ of the model fit to each of the six energy 
ranges in Table 2.  As the figure illustrates, 
the general energy dependence of the emissivity is well 
described by the electron-Bremsstrahlung $(e-B)$ (Koch \& Motz 1959; 
Fichtel et al. 1991) and nucleon-nucleon $(n-n)$ (Stecker 1989) 
production functions parameterized by Bertsch et al. (1993) 
for the solar vicinity.  The deviation from the Bertsch et 
al. production function in the 1000-10,000 MeV range has been 
seen in other studies of Galactic diffuse emission with EGRET data 
(e.g., Hunter et al. 1994; DHM; Digel et al. 1996; Hunter et 
al. 1997).  This deviation is not seen in studies of the 
isotropic emission at high latitudes (e.g., Sreekumar et al. 1998) 
and therefore is unlikely to represent a high-energy 
calibration error.  The most plausible interpretation, that 
the calculation of gamma-ray production from $n-n$ 
interactions somewhat underestimates the yield (Hunter et al. 
1997), does not affect the results presented here.

The integral gamma-ray emissivity in Orion is found to be 
$(1.65\pm0.11)\times10^{-26}$ 
s$^{-1}$ sr$^{-1}$ for $E>100$ MeV, confirming the value obtained in 
the earlier DHM analysis. 
It compares well 
with the values obtained for the Galactic plane in the solar vicinity 
in large-scale studies of diffuse emission (e.g., Lebrun \& Paul 1985;
Strong et al. 1988; Strong \& Mattox 1996), which range from
(1.54--1.8)$\times10^{-26}$ s$^{-1}$ sr$^{-1}$.
However, studies of individual clouds within $\sim500$ pc with EGRET 
data yield a wider range of integral emissivities:  
$(2.4\pm0.2)\times10^{-26}$ s$^{-1}$ sr$^{-1}$ in Ophiuchus (Hunter et al.
1994),
$(2.01\pm0.15)\times10^{-26}$ s$^{-1}$ sr$^{-1}$ in the local clouds
toward Monoceros (Digel et al. 1999), and
$(1.84\pm0.10)\times10^{-26}$ s$^{-1}$ sr$^{-1}$ in the local clouds
toward Cepheus (Digel et al. 1996).  The range of emissivities, which
we note decrease with increasing Galactocentric distance of the cloud,
suggests a fairly steep gradient of CR density at the solar circle that
is smoothed in the large-scale studies, which typically have resolutions
of 2 kpc or more.

The $X$-ratios in Table 2 are derived from the values of $A$ and $B$ for 
each energy range and are shown in Figure 4$b$.  As expected for an
intrinsic property of the molecular clouds, the value of $X$ does
not vary significantly with energy, except possibly for a decrease in
the 1000--10,000 MeV range.  The reason for the marginally-significant
decrease at the highest energies is not clear, as the highest-energy CRs 
should not be excluded from the dense, molecular parts of the clouds.

The value of $X$ derived for the $E > 100$ MeV range, 
$(1.35 \pm0.15) \times 10^{20}$ cm$^{-2}$ [K km s$^{-1}$]$^{-1}$,
is adopted here as the best overall 
value, in terms of the numbers of photons and the resolution 
of the gamma-ray observations, from our analysis. 
The non-uniformity of the exposure across the field (Fig. 1$c$) means
that this should be considered an exposure-weighted average, or more
properly an exposure-and-total column density  weighted average.  
As mentioned in \S3, the likelihood function is most sensitive to the model in 
regions with the most photons, where the exposure and gas column density 
are greatest. The exposure difference between
the Orion A and B clouds is only about 20\% (Fig. 1$c$), however,
 and the residual map
in Figure 3$c$ indicates that the same $X$-ratio applies to both clouds within
the resolution and statistics of the data.   The value of $X$ reported
in DHM, 
$(1.06 \pm0.14) \times 10^{20}$ cm$^{-2}$ [K km s$^{-1}$]$^{-1}$,
is marginally less than the value found here.  Owing to the 
greatly-improved uniformity of exposures in the dataset analyzed here, we
consider the new finding to be the more reliable.

The emissivities and $X$-ratios we find for the Orion 
region are compared in Table 3 with results from earlier 
studies.  The studies of Bloemen et al. (1984) and Houston \& 
Wolfendale (1985) were based on COS-B data, and the findings 
have been scaled here to the updated CO radiation temperature 
scale of Bronfman et al. (1988).  DHM found a value of $X$ much 
lower than that reported by Bloemen et al. (1984), and the 
lower value is confirmed here.  The instrumental background
of COS-B was significant, and had structure on the same angular
scale as the molecular clouds in Orion. The final background corrections
were not available at the time of the analysis by Bloemen et al.
(1984), and in any case small errors in the corrections for the
different COS-B viewing periods would have had a large effect on the
value of $X$ derived. 
The integral gamma-ray emissivities in Table 3 are approximately
consistent across the various studies.

The differential spectrum of the isotropic intensity 
inferred from the coefficients $C$ in Table 2 is shown in 
Figure 5.  The integrated intensity for $E > 100$ MeV is 
$(1.46 \pm 0.23) \times 10^{-5}$ cm$^{-2}$ s$^{-1}$.
The overall average 
spectrum of the isotropic emission found by Sreekumar et al. 
(1998), also shown in Figure 5, has an integral intensity of 
$(1.45 \pm 0.05) \times 10^{-5}$ cm$^{-2}$ s$^{-1}$ 
and a spectral index of $2.10 \pm 0.03$. 
On consideration of the statistical uncertainties, the intensity found
here is consistent with the expected intensity of the isotropic emission 
together with the intensity of inverse Compton 
emission and gamma-ray production on ionized hydrogen, which 
were neglected in the model (see \S 3).

\section{ Conclusions }

    This analysis of the EGRET data for the Orion region essentially
confirms the findings of the earlier work by DHM based on much less data.
No significant point sources are detected in any of four groups of viewing
periods or in the combined dataset; the marginal sources reported
by DHM are no longer even marginally significant.  The emissivity
and $X$-ratio derived for the diffuse emission are not significantly
affected if a point source is included at the position of the greatest
remaining residual intensity.
A simple linear model
for the gamma-ray emission, with adjustable parameters for the gamma-ray
emissivity, the $X$-ratio, and the isotropic intensity, including two
3EG sources just outside the field, is found to fit the
observations adequately across the EGRET energy range. 
The gamma-ray emissivity in Orion is consistent with that
found for the solar circle in large-scale studies of diffuse emission,
and its value relative to emissivities for other clouds in the solar
vicinity suggests a fairly strong gradient of CR density with
Galactocentric distance at the solar circle.  The spectrum of emissivity
is consistent with electron and proton CR
spectra approximately the same as in the solar vicinity.
The molecular
mass-calibrating $X$-ratio is 
$(1.35 \pm 0.15) \times 10^{20}$ cm$^{-2}$ (K km s$^{-1}$)$^{-1}$, and
the gamma-ray emissivity for $E >$ 100 MeV is
$(1.65 \pm 0.11) \times 10^{-26}$ s$^{-1}$ sr$^{-1}$.

\bigskip

\acknowledgements
This 
research has made use of the NASA/IPAC Extragalactic Database (NED) which is 
operated by the Jet Propulsion Laboratory, under contract with the National 
Aeronautics and Space Administration. The authors wish to thank Hans Bloemen 
for his comments on the manuscript. SWD acknowledges support from the CGRO 
Guest Investigator Program grant NAG5-2823. EA acknowledges support 
from the CGRO Guest Investigator Program grant NAG5-2872. RM acknowledges 
support from the CGRO Guest Investigator Program grant NAG5-3696.

\clearpage

\begin{figure}[t!] % fig 1
\centerline{\epsfig{file=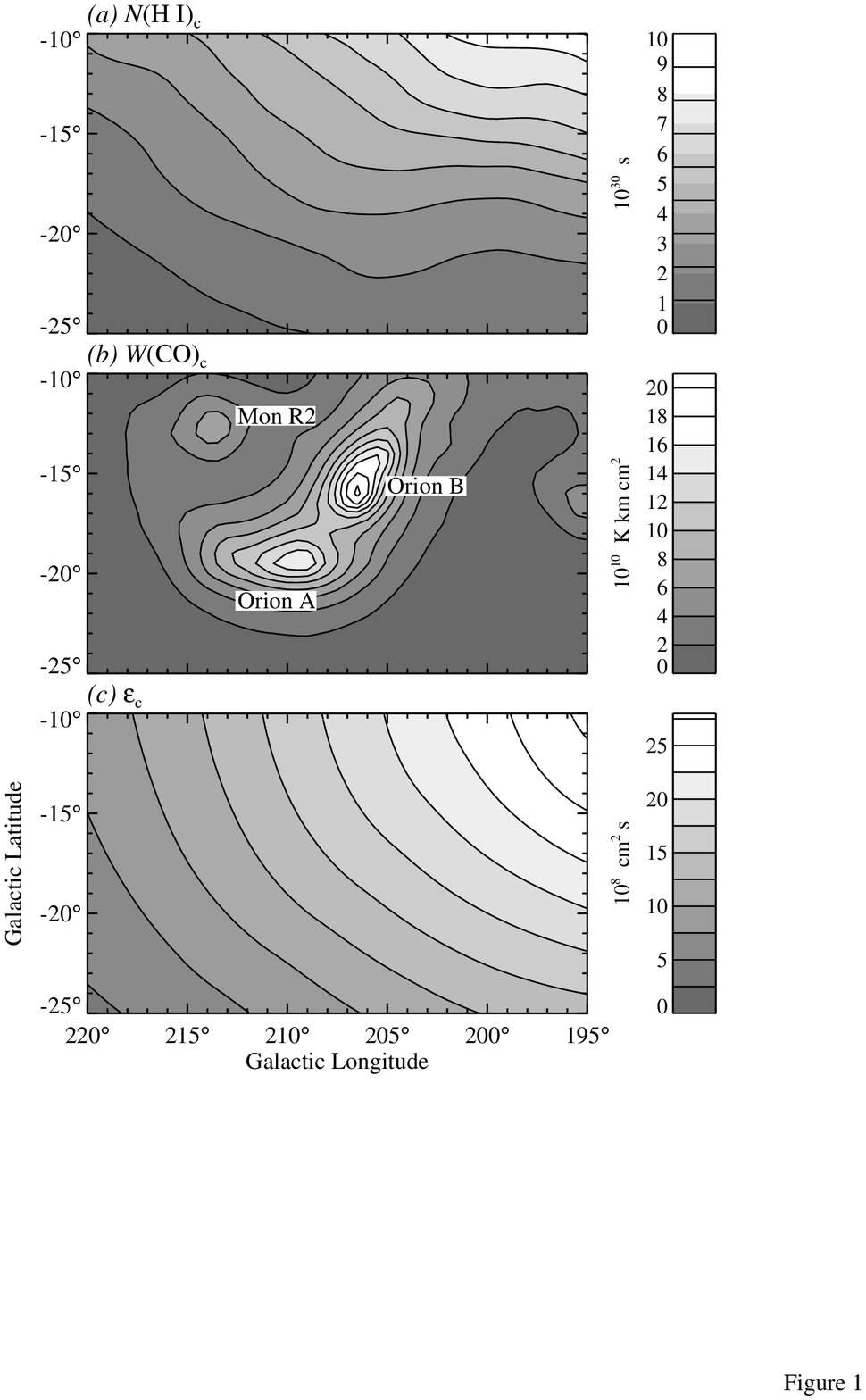,height=6.0in,bbllx=100pt,bblly=140pt,bburx=500pt,bbury=750pt,clip=.}}
\vspace{10pt}
\caption{The maps ($a$) $N{\rm (H I)}_c$, ($b$) $W{\rm (CO)}_c$,
and ($c$) $\epsilon_c$ 
described in the text, calculated for the energy range $E > 100$ 
MeV.}
\label{fig1}
\end{figure}
 
\begin{figure}[t!] % fig 2
\centerline{\epsfig{file=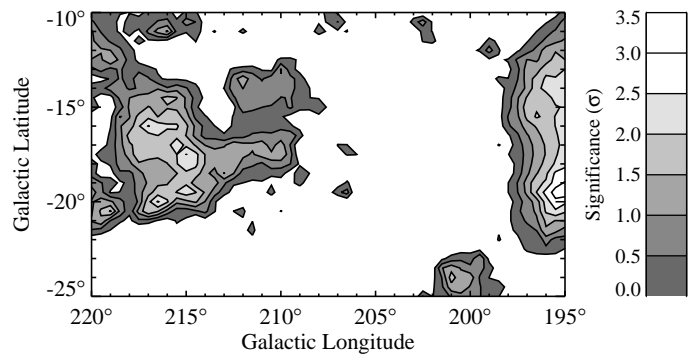,height=6.0in,bbllx=100pt,bblly=140pt,bburx=500pt,bbury=750pt,clip=.}}
\vspace{10pt}
\caption{Composite map of likelihood test statistic $TS$, the 
sum of $TS$ maps for six energy ranges spanning $E$ = 30-10,000 
MeV and representing the significance of a point source in 
the model at each position in the $30^\prime$ grid.  The contours
are in units of statistical significance in equivalent sigma, from
0.5 to 3.5 in steps of 0.5.}
\label{fig2}
\end{figure}
 
\begin{figure}[t!] % fig 3
\centerline{\epsfig{file=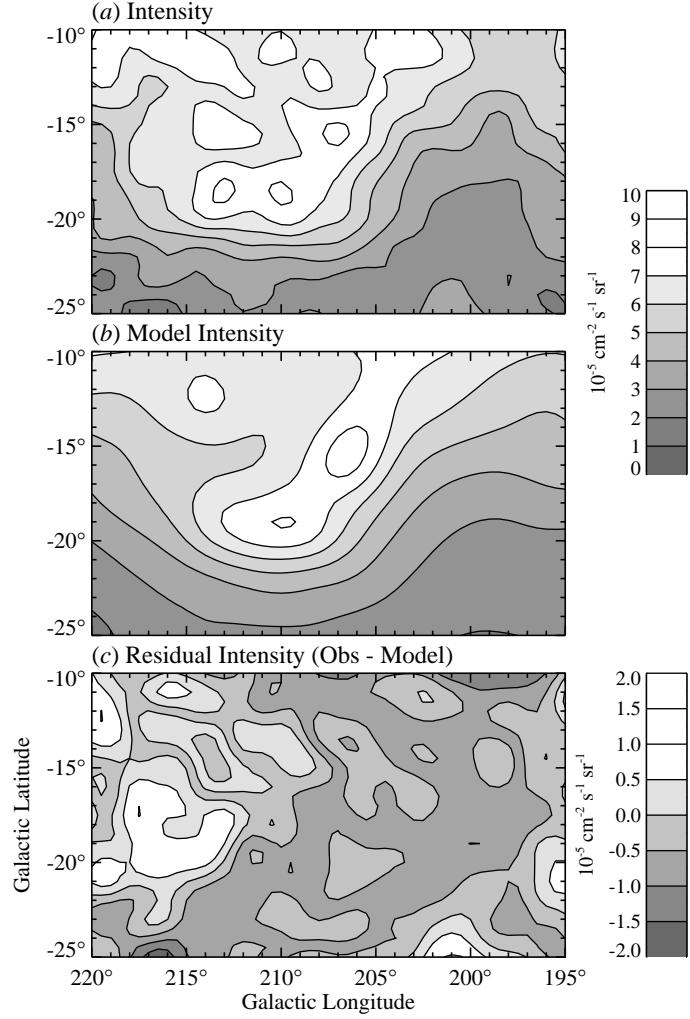,height=6.0in,bbllx=100pt,bblly=140pt,bburx=500pt,bbury=750pt,clip=.}}
\vspace{10pt}
\caption{($a$) Observed gamma-ray intensity (all viewing 
periods in Table 1), ($b$) maximum likelihood model (Eqn. 2
with the coefficients from Table 2), and ($c$) residual map 
(observed $-$ model) for the energy range $E > 100$ MeV.  The 
maps have all been smoothed slightly, by convolution with a 
2-dimensional gaussian of FWHM $1.5^\circ$, to reduce the 
statistical fluctuations in the EGRET map.}
\label{fig3}
\end{figure}
 
\begin{figure}[t!] % fig 4
\centerline{\epsfig{file=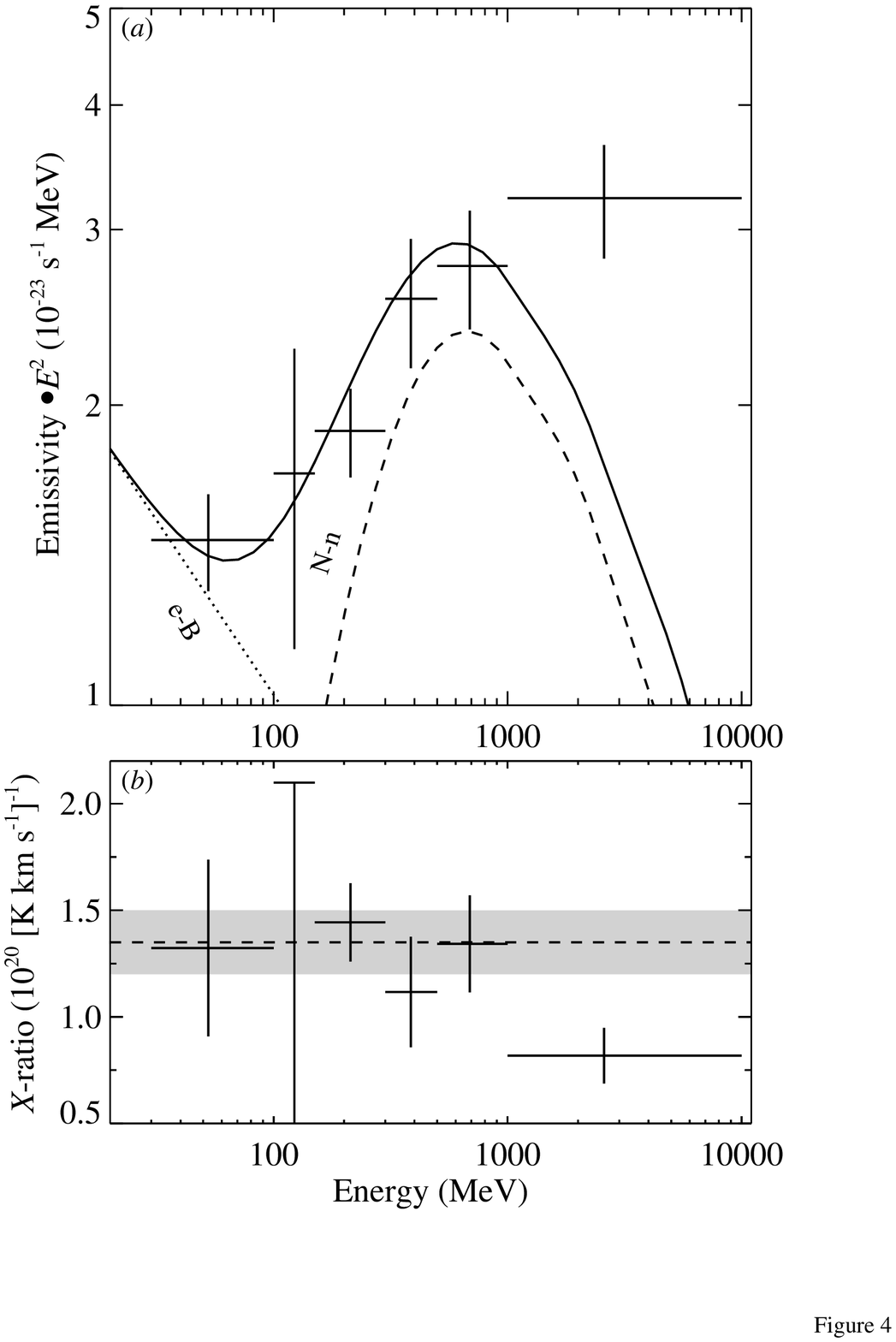,height=6.0in,bbllx=60pt,bblly=140pt,bburx=500pt,bbury=750pt,clip=.}}
\vspace{10pt}
\caption{($a$) Gamma-ray emissivity in the Orion region, 
derived from the coefficients $A$ in Table 2.  The horizontal 
bars indicate the energy ranges and the vertical bars 
the 1 $\sigma$ uncertainties of the parameters.  The solid curve 
indicates the best-fitting linear combination of the 
electron-Bremsstrahlung ($e-B$) and nucleon-nucleon ($n-n$) 
production functions used by Bertsch et al. (1993), which are 
also shown separately.  ($b$)  The energy dependence of $X = B/2A$ 
(Table 2).  For the 30-100 MeV energy range, the 2-$\sigma$ 
upper limit is shown.  The dashed line and the gray shaded 
region indicate the $X$-ratio derived for the $E > 100$ MeV range 
and its 1-$\sigma$ uncertainty.}
\label{fig4}
\end{figure}
 
\begin{figure}[t!] % fig 5
\centerline{\epsfig{file=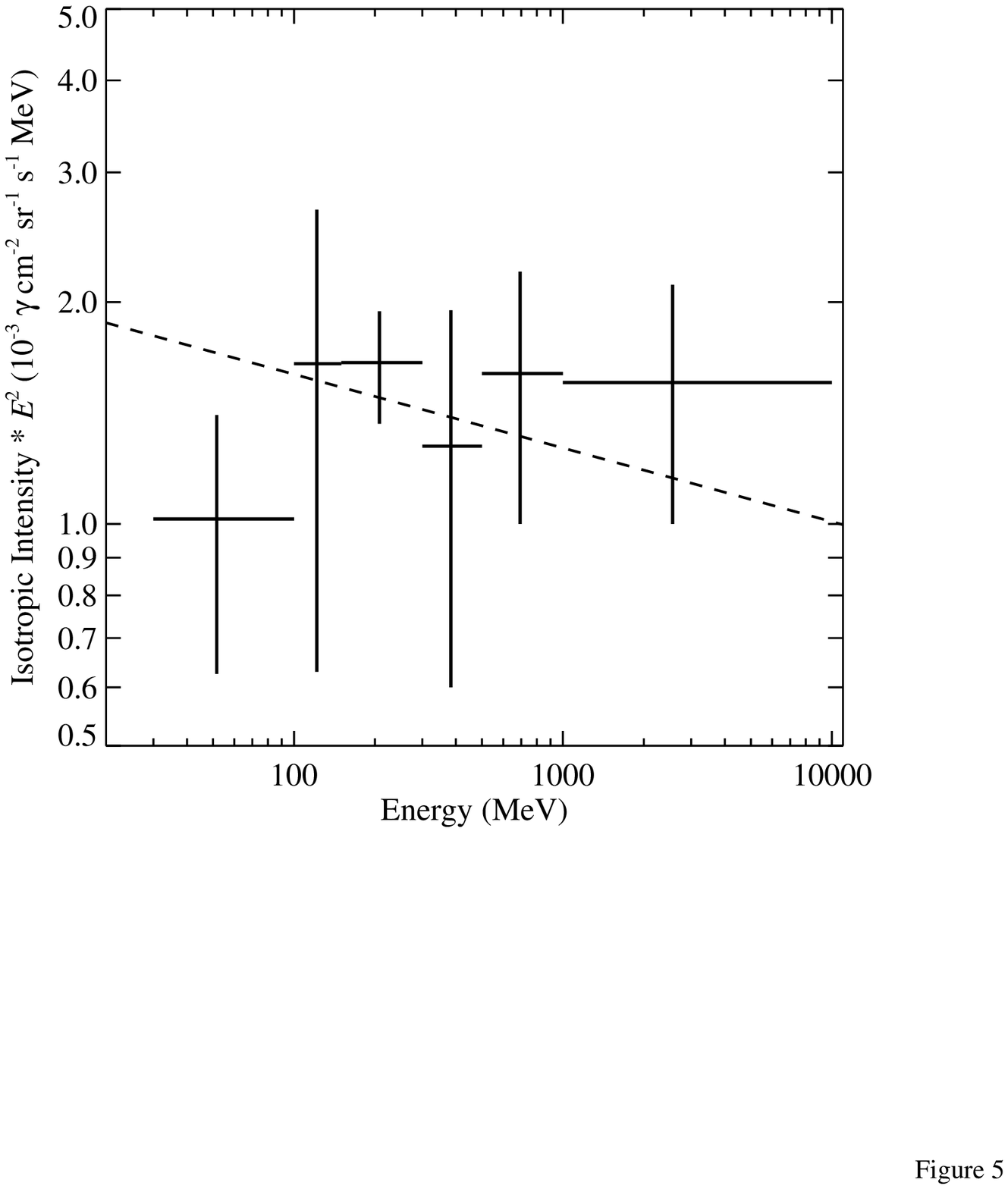,height=6.0in,bbllx=60pt,bblly=140pt,bburx=500pt,bbury=750pt,clip=.}}
\vspace{10pt}
\caption{The spectrum of the isotropic intensity toward 
Orion, derived from the coefficients $C$ in Table 2.  The 
dashed line is the overall average isotropic intensity derived by Sreekumar 
et al. (1998) for the high-latitude sky. See the text for the parameters of 
the spectra.}
\label{fig5}
\end{figure}

\end{document}